\documentclass[aps,floatfix,showpacs,twocolumn]{revtex4}
\def\beq{\begin{equation}}
\def\eeq{\end{equation}}
\def\bea{\begin{eqnarray}}
\def\eea{\end{eqnarray}}

\def\be{\begin{equation}}
\def\ee{\end{equation}}
\usepackage{color}
\usepackage{graphicx}

\begin{document}

\title{Repulsive Fermions in Optical Lattices: Phase separation versus Coexistence of Antiferromagnetism and d-Superfluidity}
\stepcounter{mpfootnote}
\author{S. Y. Chang$^{1,2,5}$}
\author{S. Pathak$^{3,4}$}
\author{N. Trivedi$^5$}
\affiliation{$^1$Institute for Theoretical Physics,
		University of Innsbruck, Technikerstr. 25, A-6020 Innsbruck Austria}
\affiliation{$^2$Institute for Quantum Optics and Quantum Information of the Austrian
              Academy of Sciences,	ICT-Geb\"aude, Technikerstr. 21a,
               A-6020 Innsbruck Austria}
\affiliation{$^3$Department of Physics, Louisiana State University, 80 Nicholson Hall, Tower Dr., Baton Rouge, LA 70803, USA}
\affiliation{$^4$Materials Research Center, Indian Institute of Science, Bangalore 560 012, India}
\affiliation{$^5$Department of Physics, The Ohio State University, 191 W. Woodruff Avenue, Columbus, Ohio 43210, USA}
\date{\today}
\begin{abstract}
We investigate a system of fermions on a two-dimensional optical square lattice in the strongly repulsive coupling regime. In this case, the interactions can be controlled by laser intensity as well as by Feshbach resonance. 
We compare the energetics of states with resonating valence bond d-wave superfluidity, 
antiferromagnetic long range order and a homogeneous state with coexistence of superfluidity and antiferromagnetism.
We show that the energy density of a hole $e_{hole}(x)$ has a minimum at doping $x=x_c$ that signals phase separation between 
the antiferromagnetic and d-wave paired superfluid phases. The energy of the phase-separated ground state is however found to be very close
to that of a homogeneous state with coexisting antiferromagnetic and superfluid orders.
We explore the dependence of the energy on the interaction strength and on the three-site hopping terms and compare with the nearest neighbor hopping {\it t-J} model.
\end{abstract}
\pacs{37.10.Jk, 74.72.-h, 74.20.-z}
\maketitle

\section{Introduction} 

The basic mechanism by which attractively interacting fermions can undergo Bardeen-Cooper-Schrieffer(BCS)-type transition through s-wave pairing is now well understood.
 The copper-oxide (CuO$_2$) materials are the first example of superconductivity arising from strong repulsive interactions~\cite{anderson87}. It is argued that  the single-band Hubbard model captures the essential physics of on-site repulsion. From variational calculations of the Hubbard model, the mechanism is now understood to be an antiferromagnetic exchange mechanism that favors singlet pairs on different sites leading to pairing in the d-wave channel~\cite{anderson04}.  These calculations show a strong deviation from the standard BCS paradigm with a separation of energy scales for pairing and long range coherence~\cite{randeria92,randeria04}. However, in the absence of a rigorous solution of the Hubbard model in two dimensions, the nature of the ground state, and in particular, whether it is a d-wave superconductor or not is still in debate~\cite{shastry10}. 

Given the difficulties of solving the Hubbard model, another route that is being currently attempted is to turn to ultracold atoms in optical lattices~\cite{jaksch98,greiner02} and use them to emulate Hubbard-like models long studied in the condensed matter physics. Unlike the materials studied in the condensed matter physics in which the interaction strengths are usually fixed,
in optical lattices the interactions can be varied over a wide range from weak to strong coupling.
 Furthermore, in the condensed matter systems changing carrier concentration usually modifies the degree of disorder, while the optical lattices are perfect and clean. 
Thus, the optical lattices allow direct simulation of one quantum system by another \cite{feynman82}.
It is hoped that a deeper understanding of superconductivity in repulsive models will suggest mechanisms for driving the transition temperature higher. At present the main bottlenecks faced in creating low entropy states in optical lattices pertain to cooling mechanisms and the role of inhomogeneity introduced by an overall trapping potential. 
Also, the multi-band effects when strongly interacting fermions are loaded in the optical lattice have to be carefully taken into account \cite{diener06,buechler10}.

In this article we discuss the problem of repulsively interacting fermions in a square optical lattice. 
The repulsion can be varied either by increasing the laser intensity or by a Feshbach resonance. For both cases we derive an effective Hamiltonian in the strong coupling regime that is independent of the interaction mechanism. 
We expect both d-wave superconducting and antiferromagnetic fluctuations to dominate in the ground state.The main question we ask is whether the ground state energetically chooses to phase separate or to stay in a homogeneous state as doping and interactions are changed. 
In the past, analogous questions were addressed in perovskite compounds such as manganites \cite{yunoki98,yunoki99,dagotto10}.
 In that case, the phase separation between undoped antiferromagnetic and hole-doped ferromagnetic regions was observed. The origin of the interaction was, however, coulombic repulsion between the localized and mobile electrons. 

Our result is that the Hubbard model does 
show a tendency to phase separate at low hole density. 
This result is based on a variational theory that compares the energetics for two possible
ground states: one that allows phase separation and another with homogeneous coexistence. 
Here, the possible biases are applied on equal degree to these different scenarios.  
Given the small difference in energy between the two solutions for phase separation and homogeneous coexistence, other factors may determine the ultimate behavior of the system:
 for example, long-range interactions may tend to favor coexistence whereas disorder may drive the system toward phase separation. Also, the effect of trapping potential needs to be better understood \cite{zhou09,mahmud11}.

\begin{figure*}[]
\includegraphics[width=4.0cm,clip]{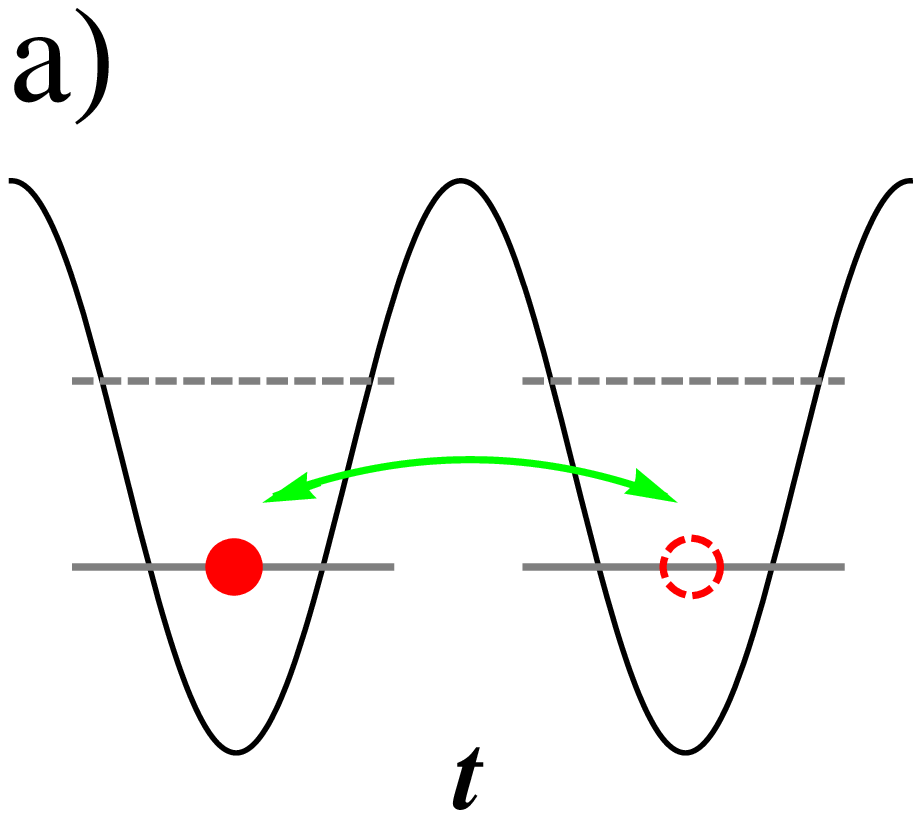}
\includegraphics[width=4.0cm,clip]{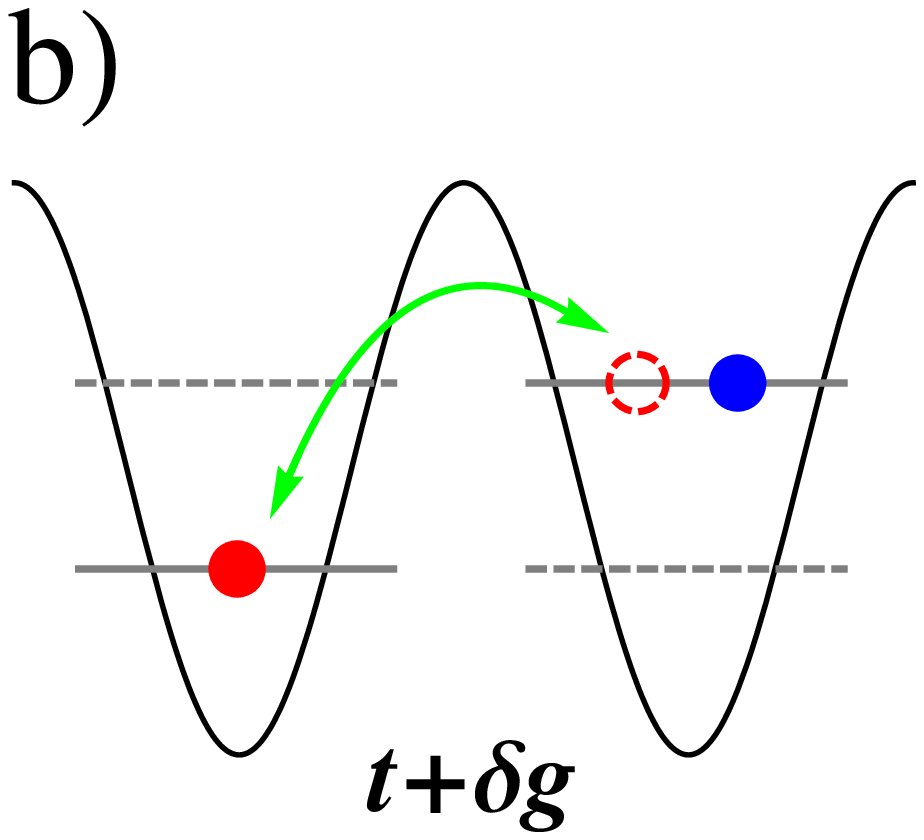}
\includegraphics[width=4.0cm,clip]{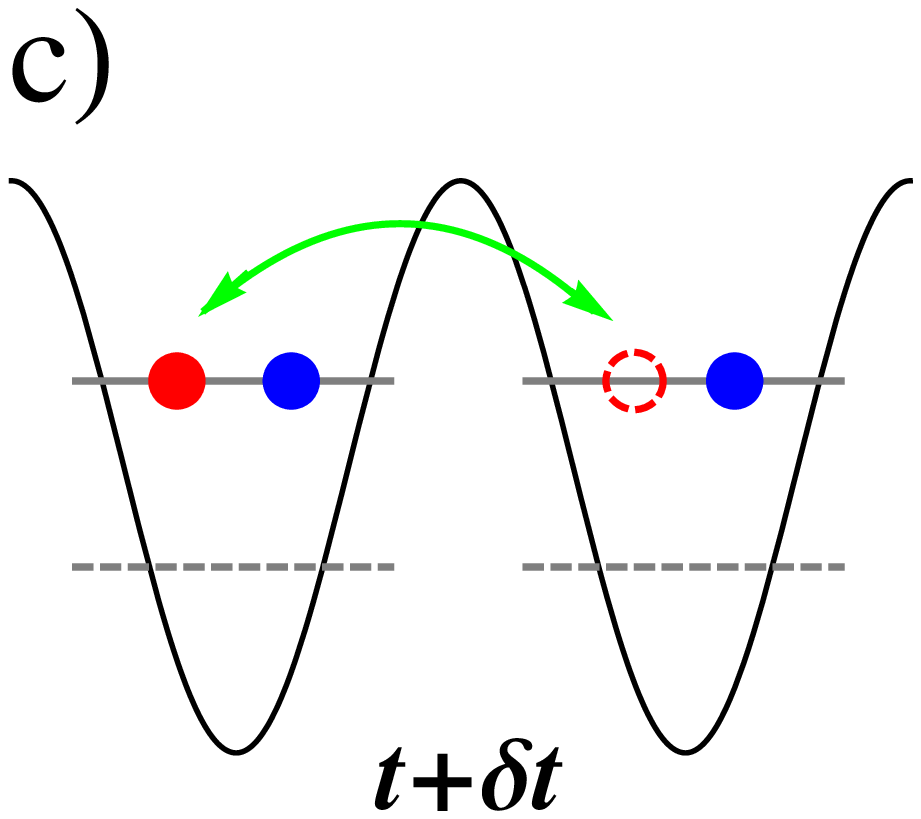}
\caption{(Color online) Here, we illustrate the effect of having a resonant two-particle state away from the single particle ground state. 
The hopping amplitudes are essentially given as the overlap of the wave functions of the neighboring sites. 
Thus, within this ``two level'' model we can identify three different hopping processes and amplitudes, a) from a single particle state to another single particle state,
 b) from a two-particle state to a single particle state or the reverse process, and c) from a two-particle state to another two-particle state.
Blue and red colors indicate fermions with opposite spins.}
\label{fig_OL}
\end{figure*}

\section{Model}
\subsection{Hamiltonian}

The interactions between fermions in the optical lattices can be controlled in two ways \cite{bloch08,schneider10}: 

i) \underline{Laser Intensity controlled}: By varying the laser intensity or the optical lattice depth $V_o$, we
can modify the tunneling rate $t$ of atoms between neighboring sites and the on-site interaction $U$. 
$t$  and $U$ can be calculated from the band structure theory.
 In the deep lattice limit, the tunneling rate is given by
$t \approx \frac{4}{\sqrt{\pi}}E_r\left(\frac{V_o}{E_r}\right)^{3/4}\exp\left[-2\left(\frac{V_o}{E_r}\right)^{1/2}\right]$ and the on-site interaction 
$U \approx\sqrt{\frac{8}{\pi}}ak_F E_r \left(\frac{V_o}{E_r} \right)^{3/4}$ \cite{bloch08}, where $a$ is the s-wave scattering length that is fixed and $k_F$ is the Fermi wave vector.
For realizing repulsive fermions, potassium isotope $^{40}$K can be used where $a$ for hyperfine states is positive ($\approx 160 a_o$ \cite{dalgarno98}) with $a_o$ defined as Bohr radius.
 Typical $a$ for $^{6}$Li is $\approx -2000a_o$. Here $E_r = \frac{h^2}{2m \lambda^2}$ is the recoil energy of the optical lattice generated by a set of counter-propagating laser beams of wavelength $\lambda$. 
 The expressions above are valid
in the non-resonant regime where $k_F a\ll 1$. It is then possible to restrict the Hilbert space to a single band Hubbard model given by
\beq
 {\cal H} = -t \sum\limits_{<i,j> \sigma} c^\dagger_{i,\sigma} c_{j,\sigma} + U \sum\limits_{i}  n_{i,+1} n_{i,-1} 
\label{eqn_hamil0}
\eeq
where $\sigma=\pm 1$ is the {\it pseudo-spin} index. The sums $\langle i,j \rangle$ run over all pairs of nearest-neighbor lattice sites
and $n_{i,\sigma} \equiv c^\dagger_{i,\sigma} c_{i,\sigma}$. We define doping parameter $x=1-N/N_{latt}$ where $N$ is the total number of fermions of both spin species and $N_{latt}$ is the number of sites in the optical lattice. 

ii) \underline{Feshbach resonance controlled}: By tuning the scattering length $a$ using a magnetic field $B$. 
Scattering length near resonance takes the form $a = a_{bg}[1-\Delta B/(B-B_o)]$ where $a_{bg}$ is the background scattering length ($\approx 160 a_o$ for $^{40}$K). Also $\Delta B \approx 8$G and $B_o \approx 202$ G for $^{40}$K.
 Here, the scattering length $a$ can be tuned from the positive values(repulsive) to zero(non-interacting) and to the negative values(attractive). 
Near the resonance, the interactions can be easily driven into a regime where they exceed the band gap or the energy difference between the lowest two levels in a single lattice site.  
 In this resonant limit the wave function of the two-particle state includes contributions from
all excited states and does not reduce to a symmetrized product of
the two single particle ground state wave functions. Thus, the minimal set of basis per site should include a single particle
ground state as well as a state with double occupancy. 
Consequently, the overlap of the wave functions for nearest neighbor sites and the
hopping amplitudes depend on the particular configuration. This situation is illustrated in the Fig. \ref{fig_OL}.
The physics of the resonant two-particle states introduces two additional parameters $\delta t$ and $\delta g$, which are incorporated into the Hamiltonian in the following form \cite{duan05,duan08}
\begin{eqnarray}
\mathcal{H} &=&-\sum\limits_{<i,j>\sigma }[t+\delta g(n_{i,\bar{\sigma}}
+n_{j,\bar{\sigma}})+\delta t(n_{i,\bar{\sigma}}n_{j,\bar{\sigma}})]c_{i,\sigma }^{\dagger }c_{j,\sigma }  \nonumber \\
&&+U_o\sum\limits_{i}n_{i,+1}n_{i,-1}.
\label{eqn_hamil1}
\end{eqnarray}
 Here, $t$ is controlled by the optical lattice depth $V_o$ as in Eq. \ref{eqn_hamil0}. 
However in the Feshbach resonance case, not just the optical lattice depth $V_o$ but also applied magnetic field and other physical parameters of the system determine the value of $U_o$. 
 Further discussions below and in the Appendix
show that within second order perturbative theory, the Hamiltonians of Eq. \ref{eqn_hamil0} and Eq. \ref{eqn_hamil1} become analogous with the identification of an {\it effective} coupling $U$.  

\subsection{Our Approach}

Given the Hamiltonian, the next step is to determine the quantum phases arising from the strongly repulsive interactions among fermions in two dimensions. At half filling, or zero doping $(x=0)$, the system is a Mott insulator with a finite gap in the charge spectrum and has long range antiferromagnetic order. As the system is doped with holes, antiferromagnetic order decreases and the system develops superfluidity described by 
a coherent superposition of the resonating valence bond (RVB) singlets \cite{anderson87}. 
Here, each singlet is formed by a pair of opposite fermion species in an orbital state with d-wave spatial symmetry. The question that arises is: What are the characteristics of the quantum phase at small hole density $x$?

In this article, we discuss two possible scenarios:

{1) \underline{Phase Separation}}:  
In this scenario, the system is macroscopically phase separated into regions of antiferromagnetism and superfluidity. We calculate the behavior of the energy density of a hole $e_{hole}(x)$ to check for phase separation. 

{2) \underline{Homogeneous}}: Here, we consider a single homogeneous phase with coexisting antiferromagnetic
and superfluid orders occupying the whole system. We evaluate the energetics of variational states that
include both superfluid and antiferromagnetic orders.

By comparing with the more restricted standard {\it t-J} model we find that the presence of effective 3-site hopping terms in the  {\it full} Hamiltonian defined in Eq. \ref{eqn_heff} leads to a lower energy at non-zero hole density $x$. We also find a stronger tendency to phase separate in the {\it full} Hubbard Hamiltonian compared to the more restrictive {\it t-J} model.

\subsection{Effective Hamiltonian in Strong Coupling} 

In the large $U/t$ limit, the Hamiltonian in Eq. \ref{eqn_hamil0} can be transformed into a block diagonal form where the blocks preserve
the total number of the doubly occupied sites and the different blocks are connected by the kinetic/hopping operators (see Appendix). 
The effective Hamiltonian ${\cal H}_{eff}$ in the strong coupling limit is obtained by a canonical transformation\cite{macdonald88,cohen}
\beq
{\cal H}_{eff} = {\cal V} + {\cal K}_0 + \frac{1}{U} [{\cal K}_1, {\cal K}_{-1}] + {\cal O}(1/U^2)
\label{eqn_eff0}
\eeq
 where the kinetic operator ${\cal K}_m$ changes the number of  doubly occupied sites by an integer number $m$.
${\cal V}$ is the interaction term. The term given by a commutator (second order term) in Eq. \ref{eqn_eff0} contains two consecutive hoppings. Since we consider $U/t >>0$, we impose the non-double occupancy constraint at each site in order to obtain the low energy Hamiltonian: 
${\cal H}_{eff} \rightarrow {\cal P}_G {\cal H}_{eff} {\cal P}_G$  where
${\cal P}_G = \prod\limits_{i}(1-  n_{i,1}  n_{i,2})$  is the {\it full} Gutzwiller projector.
As a result, the interaction term $\cal V$ is projected out and the Hamiltonian can be cast as
\begin{widetext}
\beq
{\cal H}_{eff}  =  -t \sum\limits_{<i,j>\sigma}  f^\dagger_{i,\sigma} f_{j,\sigma}
  -\frac{t^2}{U} \sum\limits_{<i',i>\sigma'; <i,j>\sigma} f^\dagger_{i',\sigma'} d_{i,\sigma'} d^\dagger_{i,\sigma} f_{j,\sigma} 
\label{eqn_heff}
\eeq
\end{widetext}
The summations run over all independent indices and Eq. \ref{eqn_heff} is what we call the projected or {\it full} Hamiltonian to distinguish it from the more restricted {\it t-J} model.
In the second sum, we can identify two-site (when $i' = j$ ) and three-site (when $i' \ne j$) processes. 
 In the {\it t-J} model, an additional restriction $i' = j$ is imposed such that only two-site
exchange is allowed. In Figs. \ref{fig_1b} and \ref{fig_2b}, we show the effects of this constraint. 
We have introduced definitions: $f^\dagger_{i,\sigma} \equiv h_{i,\bar{\sigma}}c^\dagger_{i,\sigma} $ 
the creation operator of a singly occupied site  with spin $\sigma$ particle and $d^\dagger_{i,\sigma} \equiv c^\dagger_{i,\sigma}h_{i,\sigma}n_{i,\bar{\sigma}}$
the creation operator of a doubly occupied site  by adding a particle of spin $\sigma$ to a singly occupied site with a spin $\bar{\sigma}$
particle where $h_{i,\sigma} \equiv 1- n_{i,\sigma}$. Rewriting the Hamiltonian in terms of the {\it pseudo-particle} operators $f_{i,\sigma}$, $f^\dagger_{i,\sigma}$, $d_{i,\sigma}$ and $d^\dagger_{i,\sigma}$
 makes clear the permitted physical processes. However, these operators do not satisfy the usual anti-commutator nor commutator relations and straightforward theoretical treatment is difficult. Thus, we rely on the variational formalism whose accuracy depends on the chosen ansatz \cite{edegger07,anderson06,sorella02}. In the following section, we discuss the different variational wave functions.\\

 For the fermions near resonance (Eq. \ref{eqn_hamil1}), the same perturbative Hamiltonian (Eq. \ref{eqn_heff}) is derived with an {\it effective} interaction $U$ which has two contributions 
\beq
U= \frac{U_o}{(1+\delta g/t )^2}.
\label{eqn_strength}
\eeq
Here, $U_o$ is controlled by the optical lattice depth and the magnetic field while $\delta g/t$ parameter accounts for the appearance of a two-particle state that goes beyond the single band model (see Appendix).
 The $\delta g$ dependence comes from the allowed virtual processes while the $\delta t$ contribution does not appear in the 
second order perturbation. As a consequence, we now consider the phase diagram of the Hamiltonian in Eq. \ref{eqn_heff} as a function of $U/t$ and doping $x$, regardless of the different ways of tuning the interaction $U$.

\subsection{Variational Wave functions}
 
{\bf 1) Phase separated solution using Maxwell construction:}
Phase separation results from thermodynamic considerations that involve two pure phases. In the following, we give their description leaving the details of Maxwell construction to the section \ref{sec_dr} (Discussion of Results).  

The pure superfluid variational state is described by  \cite{randeria04}
\bea
|\Psi_{SF_d}\rangle  & = & {\cal J} {\cal P}_G {\cal P}_N |SF_d \rangle_{MF} \nonumber \\
  & = & {\cal J} {\cal P}_G \left[ \sum\limits_{\bf k} \alpha_{\bf k} c^\dagger_{{\bf k},\sigma}c^\dagger_{-{\bf k},\bar{\sigma}} \right]^{N/2} |0 \rangle \nonumber \\ 
\label{eqn_rvb}
\eea
$|\Psi_{SF_d}\rangle$ is also called the RVB wave function when Jastrow correlation is set to unity. But as we discuss below, $|\Psi_{SF_d}\rangle$
is more general and can also describe a system with long range antiferromagnetism for an appropriate choice of the Jastrow correlations.
 $|SF_d \rangle_{MF}$ is the solution of the simple mean field Hamiltonian of the form:
${\cal H}_{MF}  = \sum_{{\bf k}\sigma} \xi_{\bf k} c^\dagger_{{\bf k},\sigma} c_{{\bf k},\sigma} 
+ \sum_{\bf k} \Delta_{\bf k} [c^\dagger_{{\bf k},\uparrow} c^\dagger_{-{\bf k},\downarrow} + h.c.] $. 
Thus, the parameters $\alpha_{\bf k}$ are given as $ \alpha_{\bf k} =  \Delta_{\bf k}/\left[\xi_{\bf k} +  \sqrt{\xi_{\bf k}^2 + \Delta_{\bf k}^2)}\right]$.
The d-wave parametrization $\Delta_{\bf k}=\Delta_d(\cos k_x -\cos k_y)$ is always energetically favorable in comparison
with the s-wave symmetric parametrization $\Delta_{\bf k} = \Delta_s$. We get the appropriate $\mu_{var}$  from the number equation 
$N  =  \sum\limits_{\bf k} \left[ 1-\xi_{\bf k}/\sqrt{\xi_{\bf k}^2+\Delta_{\bf k}^2} \right]$. 
Independent optimization of $\mu_{var}$ has a negligible effect on the energy. 
For our fully projected wave function of Eq. \ref{eqn_rvb}, we avoid attaching physical meaning to the
parameters $\Delta_d$ and $\mu_{var}$. They are taken simply as two independent degrees of freedom in the variational wave function.
In order to characterize the quantum phases, the appropriate order parameters that characterize the SF and AF long range order are calculated explicitly
in the optimized wave functions.
It can be seen that for the wave function of Ref. \cite{pathak08}, when long range hopping parameters $t'$ and $t''$ of $\xi_{\bf k}$ are set to zero, the $\alpha_{\bf k}$
parametrization of the wave function becomes equivalent to that of Ref. \cite{giamarchi90,himeda99} and  Eq. \ref{eqn_rvb}.
 Unlike the case of high $T_c$ materials where the $t'$ and $t''$ terms are important to describe the correct band structure of the materials, in the context of cold atoms in optical lattices, these extra parameters can be set to zero. 

The quantum phase at half filling (zero hole doping) has long range antiferromagnetic (AF) order \cite{trivedi90}.
For large enough system sizes, a broken SU(2) symmetric ground state becomes
likely. In a ground state that has antiferromagnetic long range order, we note that even at zero doping, the singlet correlation accounts for large part of the energy
contribution \cite{liang88}. Thus, we keep the singlet correlation mechanism of Eq. \ref{eqn_rvb}  and
adjust the fluctuations around the Neel state with a Jastrow factor that explicitly breaks the SU(2) symmetry:
${\cal J} = e^{ m_{af} \Delta_{af}}$. Here, 
\beq
m_{af} = \frac{2}{N_{latt}}\left[ \sum\limits_{i \in L_1} S_i^z - \sum\limits_{i \in L_2} S_i^z \right]
\label{eqn_maf}
\eeq
is the staggered magnetization operator with  $S_i^z = 1/2(n_{i,+1}-n_{i,-1})$. 
Also,  $L_j$ is the sub-lattice of index $j$. $\Delta_{af}$ can be interpreted as a local magnetic field that
breaks the SU(2) symmetry. Thus, for the zero doping case we are assuming three variational parameters: $\Delta_d$, $\Delta_{af}$,
and $\mu_{var}$. The improvement in energy due to this choice of the Jastrow factor is $\sim 2\%$ which is much larger than the typical statistical errors. It was found that the optimum value $\Delta_{af} = 0.35$ not only produces lower energy but also the correct staggered magnetization \cite{trivedi90}.

 {\bf 2) Homogeneous solution with coexistence of antiferromagnetism and superfluidity:}
A homogeneous state with both the d-wave superfluid and staggered magnetic long range order(denoted by $|AF+SF_d \rangle$) is based on the mean field solution \cite{giamarchi90,himeda99,lee97,lee03,pathak08}. 
This state is favorable close to the zero hole doping $x = 0$ up to a
critical $x = x_c$. While for $x_c \le x \le x_c'$, a pure d-wave superfluid (SF$_d$) phase is favored.
The initial mean field $|AF+SF_d\rangle_{MF}$ state is obtained by solving the Hamiltonian
\begin{widetext}
\bea
{\cal H}_{MF} & =& \sum_{{\bf k}\sigma} \xi_{\bf k} [c^\dagger_{{\bf k},\sigma} c_{{\bf k},\sigma} +
\xi_{{\bf k} + {\bf Q} } c^\dagger_{{\bf k}+{\bf Q},\sigma} c_{{\bf k}+{\bf Q},\sigma}  ]
  +  \sum_{\bf k} \lbrace { \Delta_{\bf k} [c^\dagger_{{\bf k},1} c^\dagger_{-{\bf k},2} + h.c.] +
\Delta_{{\bf k}+{\bf Q}} [c^\dagger_{{\bf k}+{\bf Q},1} c^\dagger_{-({\bf k}+{\bf Q}),2} + h.c.]  \rbrace} \nonumber \\
& + &  \frac{1}{4} \sum_{{\bf k}\sigma} \sigma m_N({\bf Q}) [c^\dagger_{{\bf k}+{\bf Q},\sigma} c_{{\bf k},\sigma} + c^\dagger_{{\bf k},\sigma} c_{{\bf k}+{\bf Q},\sigma}]
\eea
\end{widetext}
where the $\bf k$ vector is restricted to the {\it reduced Brillouin zone} and ${\bf Q} = (\pi,\pi)$. The diagonalization produces two spin
density wave bands denoted by subindices $a$ and $b$. The dispersion relations are given as 
\bea
E_{a} ({\bf k}) & = & \frac{\xi_{\bf k} + \xi_{{\bf k} + {\bf Q}} + \sqrt{(\xi_{\bf k} - \xi_{{\bf k} - {\bf Q}})^2+m_N({\bf Q})^2}}{2} \nonumber \\
E_{b} ({\bf k}) & = & \frac{\xi_{\bf k} + \xi_{{\bf k} + {\bf Q}} - \sqrt{(\xi_{\bf k} - \xi_{{\bf k} - {\bf Q}})^2+m_N({\bf Q})^2}}{2} 
\eea
with $\xi_{\bf k} = -2 t (cos(k_x) + cos(k_y))-\mu_{var}$ and $m_N({\bf Q}) = $ magnetic variational parameter. The corresponding quasiparticle operators are
\bea
d_{{\bf k},a,\sigma} & = & \alpha_{\bf k} c_{{\bf k},\sigma} + \sigma \beta_{\bf k} c_{{\bf k} + {\bf Q},\sigma} \nonumber \\
d_{{\bf k},b,\sigma} & = & -\sigma \beta_{\bf k} c_{{\bf k},\sigma} +  \alpha_{\bf k} c_{{\bf k} + {\bf Q},\sigma}.
\eea
From these, the ground state ansatz can be written as
\begin{widetext}
\beq
|\Psi_{AF + SF_d}\rangle   =  {\cal J} {\cal P}_G {\cal P}_N |AF+SF_d \rangle_{MF}  
   =    {\cal J} {\cal P}_G \left[ \sum\limits_{\bf k} A_{\bf k} d^\dagger_{{\bf k},a,+1 }d^\dagger_{-{\bf k},a,-1} +
B_{\bf k} d^\dagger_{{\bf k},b,+1 }d^\dagger_{-{\bf k},b,-1} \right]^{N/2} |0 \rangle. 
\label{eqn_rvb+af}
\eeq
\end{widetext}
Here, $A_{\bf k} =  \Delta_{\bf k}/\left[E_a({\bf k}) +  \sqrt{(E_a({\bf k}))^2 + \Delta_{\bf k}^2)}\right]$
and  $B_{\bf k} =  -\Delta_{\bf k}/\left[E_b({\bf k}) +  \sqrt{(E_b({\bf k}))^2 + \Delta_{\bf k}^2)}\right]$
with   d-wave symmetric parametrization $\Delta_{\bf k} = \Delta_d(cos(k_x)-cos(k_y))/2$.
${\cal P}_N$ projects the wave function into the subspace of total number of particles $N= N_1+N_2$ 
with equal populations $N_1=N_2$.
For this variational wave function, the optimization parameters are $\Delta_d$, $m_N({\bf Q})$, and  $\mu_{var}$.

In the hole doped regime, the contribution from particle-particle
correlations is small and the hole-hole correlations dominate \cite{sorella02}. We assume a Jastrow factor $\cal J$
of the form ${\cal J} = e^{1/2\sum_{i,j} h_i h_j v_{ij}}$ where the hole density at {\it i-th} site is $h_i \equiv h_{i,1} h_{i,2}$ and the correlation
function is parametrized by $v_{ij} = v/|{\bf r}_i - {\bf r}_j|$. We find an improvement in energy of the 
order $\sim 0.1 \%$ with $v<0$. Here, the distance between sites $r_{ij} = |{\bf r}_i - {\bf r}_j|$ must be between the closest images across the tilted boundaries as we later discuss.

\section{DISCUSSION OF RESULTS}
\label{sec_dr}

 In analogy with the high $T_c$ materials, we consider $U$ values in the Eq. \ref{eqn_heff} centered at $13.33$
(in units of $t$). For the case of the simple single band model (Eq. \ref{eqn_hamil0}) this $U$ value is solely controlled by the optical lattice depth $V_o$ while for the model of Eq. \ref{eqn_hamil1}, $U$ is
controlled by $V_o$ as well as by Feshbach resonance(magnetic field). For our discussion, we assume the parametrization of Eq. \ref{eqn_strength}
with $U_o = 13.33$ and $\delta g/t$ varied. For the non-resonant case, this is just a way of getting different values of $U$, while for the 
resonant case $\delta g/t$ has physical meaning. 

 The quantum expectation values are evaluated by Monte Carlo multi-dimensional integration. 
 In order to avoid numerical instabilities along the $k_x = k_y$ points,
we take a patch of lattice sites with the tilted boundary conditions as our system frame \cite{kaxiras88,randeria04}.
 The size of the system is $N_{latt} = L^2+1$ sites, where $L$ is an odd number. In this way, we also avoid the frustration
 of the antiferromagnetic phase (one particle of spin +1(-1) should be surrounded by
4 particles of spin -1(+1)) at the half filling and low hole doping limits.
The pairing wave function in real space is obtained by Fourier transforming $\phi_k=u_k/v_k$.
This function is continuous when crossing the tilted boundaries. 

The injection of holes to the antiferromagnetic(AF) phase at half filling destabilizes the AF and beyond a critical hole doping density $x_c$, the AF order is completely destroyed giving
way to the homogeneous d-wave superfluid (SF$_d$) (Eq. \ref{eqn_rvb}). There are two possible scenarios
of AF $\leftrightarrow$ SF$_d$ transitions considered here: 
1) a first order phase transition between AF and SF$_d$ phases as a function of doping.
 2) a second order phase transition from a homogeneous $AF + SF_d$ phase at low doping densities to the $SF_d$ phase. The energetics of this scenario is calculated using the wave function of Eq. \ref{eqn_rvb+af}.
In both cases, the pure SF$_d$ phase exists in the hole doping regime $x_c < x < x_c'$. For $x > x_c'$, the system becomes a normal Fermi fluid (NFF) phase with none of the long range antiferromagnetic and superfluid correlations.
In all cases, the s-wave symmetric superfluidity is energetically disfavored.

\subsection{Magnetic and Superfluid order parameters}

\begin{figure}[]
\includegraphics[angle=0,width= 8cm,clip]{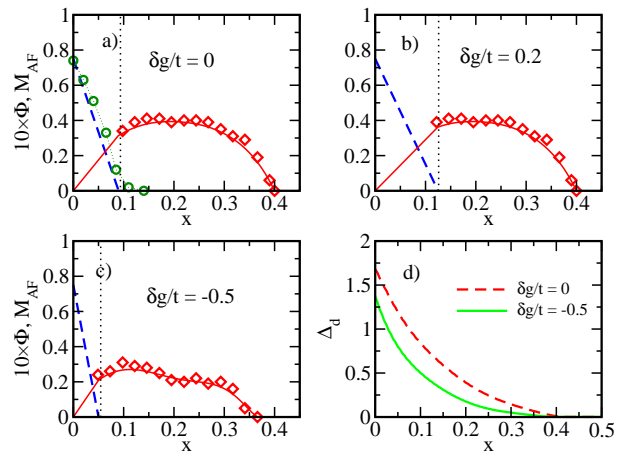}
\caption{(Color online) Superfluid long range order $\Phi$ (diamonds, continuous line) and antiferromagnetic order parameter $M_{AF}$ (dashed line) at different
hole densities and coupling $U$ (plots a-c). These figures correspond to the scenario with phase separation. 
Here, we assumed the {\it full} Hamiltonian of Eq. \ref{eqn_heff} and the interaction as parametrized by
$U = U_o/(1+\delta g/t)^2$ with $U_o =13.33$ in units of $t$ (Eq. \ref{eqn_strength}). 
In the phase separated region $0 < x < x_c$ (horizontal dotted line), the order parameters are assumed to depend linearly on $x$, 
so that $\Phi(x) = \Phi(x_c) x/x_c$ and $M_{AF}(x) = M_{AF}(0)\left[1-x/x_c \right]$. For comparison, we also show  $M_{AF}(x)$ for the homogeneous phase scenario(circles) at the interaction given by $\delta g/t = 0$ (plot a). 
We find that the order parameters for the phase separation scenario and the homogeneous phase scenario are very close
to each other. $U=9.26 < U_o$ ($\delta g/t = 0.2$, plot b) is more favorable
for superfluidity with a larger region of phase coexistence, 
while the opposite is true for $U = 53.32 > U_o$ ($\delta g/t = -0.5$, plot c).
There is a distinguishable difference for the optimizing superfluid parameter $\Delta_d$ between $\delta g/t = 0$ and  $\delta g/t = -0.5$ (plot d).
However, there is only a small difference for $\Delta_d$ between $\delta g/t = 0$ and
$\delta g/t = 0.2$ cases; thus we use the same parametrization for both. 
The optimum value of the antiferromagnetic parameter  $\Delta_{af}$ was $\sim 0.35$ at $x=0$. It has a rather weak dependence on $\delta g/t$ (that is, on $U$). } 
\label{fig_4}
\end{figure}

 In order to characterize the quantum phases, we calculate the following long range orders; d-wave order  
$\Phi  = \lim\limits_{j \rightarrow \infty}  \phi(j)$. Here, we defined $\phi(j) \equiv \sqrt{2/N \langle \sum\limits_{i} B^\dagger_{i+j} B_{i} \rangle} $ with 
 $ B^\dagger_i \equiv 1/2 (c^\dagger_{i,+1} c^\dagger_{i+\hat{x},-1} - c^\dagger_{i,-1} c^\dagger_{i+\hat{x},+1})$.
And the long range order for the staggered magnetization $M_{AF} = \langle m_{af} \rangle$ (Eq. \ref{eqn_maf}).  
 As shown in the Fig. \ref{fig_4},
$\Phi$ has a region of favorable d-wave paring at finite hole densities. At zero hole density, we found that the broken SU(2) symmetry wave function produces the correct finite size behavior of $M_{AF} = 0.75(2)$ in agreement with previous variational calculations \cite{trivedi90,pathak08} for the considered system sizes ($N_{latt} = 50$ to $226$). These long range orders, however, cannot
provide information on the nature of the phase transition (first or second order). In fact, they are shown to be in close
agreement for the phase separation and the homogeneous phase cases (Fig. \ref{fig_4}). 
We find that for $U > U_o$ where the exchange is suppressed, 
 the superfluid pairing is also suppressed (see Fig. \ref{fig_4}), while in the $U < U_o$ case the relative mobility $t/U$
of the particles is enhanced and it becomes favorable for d-wave superfluidity. 

\subsection{Phase separation vs Coexistence at low hole doping}
\begin{figure}[]
\includegraphics[angle=0,width=6cm,clip]{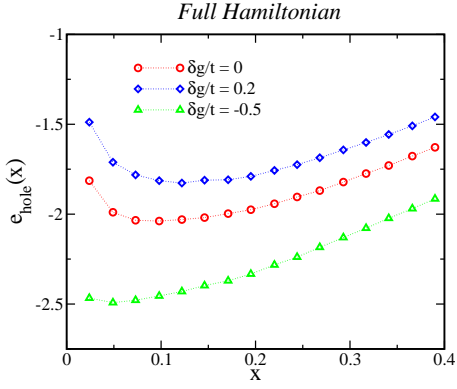}
\caption{(Color online) Energy density per hole $e_{hole}(x)$ as a function of the hole density $x$ is shown for the 
{\it full} Hamiltonian model. The interaction strengths corresponding to $\delta g/t = -0.5$ ($U/t = 53.32$), $0$ ($U/t = U_o/t = 13.33$) and $0.2$ ($U/t = 9.26$)
 cases are considered. We observe
a minimum of $e_{hole}(x)$ at $x = x_c > 0$ signaling phase separation in the region $0<x < x_c$. 
$x_c$ increases monotonically with decreasing $U$ while still
remaining in the large $U$ regime.
The results are obtained on systems of $N_{latt} = 9^2+1=82$ lattice sites.
The unit of energy is $t$.} 
\label{fig_1}
\end{figure} 

In order to characterize the inhomogeneous mixture phase, the thermodynamic considerations are as following:
From thermodynamic constraints, the energy per site of the ground state ${\cal E}(x) = E(x)/N_{latt}$ has to be a convex function 
$\partial^2 {\cal E}(x)/\partial x^2 \ge 0$. The range of $x \in (x_0,x_1)$ where 
$\partial {\cal E}(x)/\partial x = constant$ (or $\partial^2 {\cal E}(x)/\partial x^2 =0 $) implies that the number of holes can be varied while the
chemical potential $\mu = \partial {\cal E}(x)/\partial x$ is kept the same. This is a signature of the
first order phase transition.  In this regime, the system is a mixture of phase $I$ at $x_0$ and phase $II$ at $x_1$.
In our case, $x_0 = 0$ and phase $I$ has antiferromagnetic order(AF) while $x_c \equiv x_1$ and phase
$II$ is the d-wave superfluid(SF$_d$). In order to check whether there is an interval defined by $x=0$ and $x_c > 0$, we
should check for the {\it flatness} of ${\cal E}(x)$ \cite{boninsegni08}. However, because {\it we lack the knowledge of the exact mixed phase 
ground state at all $x$} we use the Maxwell construction: we calculate the energy per hole $e_h(x)$ and check for the existence of the 
minimum\cite{emery90} of
\beq
e_h(x) = \frac{{\cal E}_{II}(x) - {\cal E}_I(0)}{x}, 
\eeq 
where ${\cal E}_{I,II}(x)$ are the energy densities of the pure phases (AF and SF$_d$).
This is equivalent to finding $x_c$ such that $\frac{\partial {\cal E}_{II}(x)}{\partial x}|_{x_c} =  \frac{{\cal E}_{II}(x_c) - {\cal E}_I(0)}{x_c}$.
For all cases considered here,  $e_h(x)$  has a minimum (Fig. \ref{fig_1}) and Maxwell construction with non zero $x_c$ is possible.
 In order to avoid pair breaking effects, we introduce two holes at a time by removing one particle
of each species. From Fig. \ref{fig_1}, we find that the critical doping strengths are:
for $\delta g/t = 0$, $x_c \approx 0.09$, for $\delta g/t = 0.2$, $x_c \approx 0.12$, and for $\delta g/t = -0.5$, $x_c \sim 0.05 $. Correspondingly, the long range orders within the
region $0< x < x_c$ are proportional to the areas of the pure quantum phases; that is, the AF phase with $x=0$
and SF$_d$ phase with $x=x_c$ (see the caption of Fig. \ref{fig_4}).

\subsection{Ground State Energies}
\begin{figure}[]
\includegraphics[angle=0,width=6cm,clip]{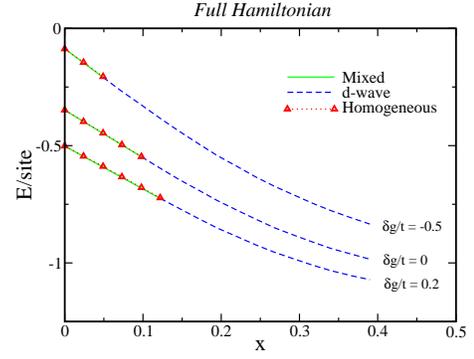}
\caption{(Color online) The ground state energies of Hamiltonian in Eq. \ref{eqn_heff} as a function of doping $x$ 
for various interaction strengths parametrized by $\delta g/t$:
$\delta g/t = -0.5$ ($U/t = 53.32$), $0$ ($U/t = U_o/t = 13.33$) and $0.2$ ($U/t = 9.26$).
 Within the region defined by $x \in(0,x_c)$, we also show the energies corresponding to the first order (continuous line)
and the second order (triangle) phase transition scenarios. 
The energies are shown to be degenerate (within $ < 0.01\%$ error bars)
for the mixed and homogeneous cases. SF$_d$ smoothly transitions to the normal Fermi fluid NFF at $x \gtrsim x_c'$ ($x_c' \approx 0.4$) when $\Delta_d \rightarrow  0$.The unit of energy is $t$. } 
\label{fig_2}
\end{figure}

\begin{figure}[]
\includegraphics[angle=0,width=5cm,clip]{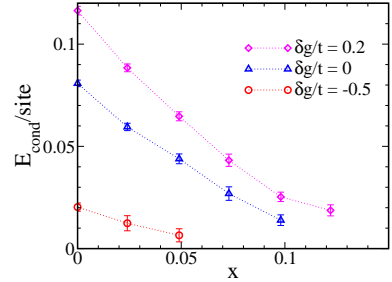}
\caption{(Color online) Condensation energy per site at different values of $\delta g/t$ for the 
Hamiltonian of Eq. \ref{eqn_heff}. The unit of energy is $t$. Hole densities $x$ between $0$ and $x_c$ are considered. In this case, $x=0$ corresponds to the pure antiferromagnet while $x=x_c$ corresponds to the pure superfluid phase.
 For $0 < x < x_c$, the condensation energy is that of the homogeneous solution.} 
\label{fig_2c}
\end{figure}

 The ground state energies for the {\it full} Hamiltonian are shown in the Fig. \ref{fig_2}. We notice that the mixed phase and the homogeneous phase energies are closely
degenerate within the region of $x \in(0,x_c)$ in all of the considered cases.  
The kinetic energy has no contribution at zero doping. In this case, the only contribution comes from exchange. As holes are injected to the
system, exchange of particles becomes suppressed while the kinetic contribution increases. In the normal phase regime beyond $x_c^\prime$, most of the contribution to the energy comes from the
kinetic term. The composition of energies for the {\it t-J} model is shown in the Fig. \ref{fig_3}.

 From the Fig. \ref{fig_2c}, we see a monotonically decreasing behavior of the condensation energy $E_{cond}$ which is the difference in energies between the normal state and the state with quantum coherence.
 Since $E_{cond}$ is at its largest for the pure antiferromagnetic phase, we can argue that once the
system is phase separated it would not transition into the homogeneous phase. 
This would be true even when the superfluid portion of the system is destroyed by thermal fluctuations as long as the antiferromagnetic portion retains its phase coherence. Thus, we argue that the phase separation scenario
is more {\it robust} than the homogeneous coexistence.

\subsection{Comparison of the {\it Full} vs {\it t-J} models}
\begin{figure*}[]
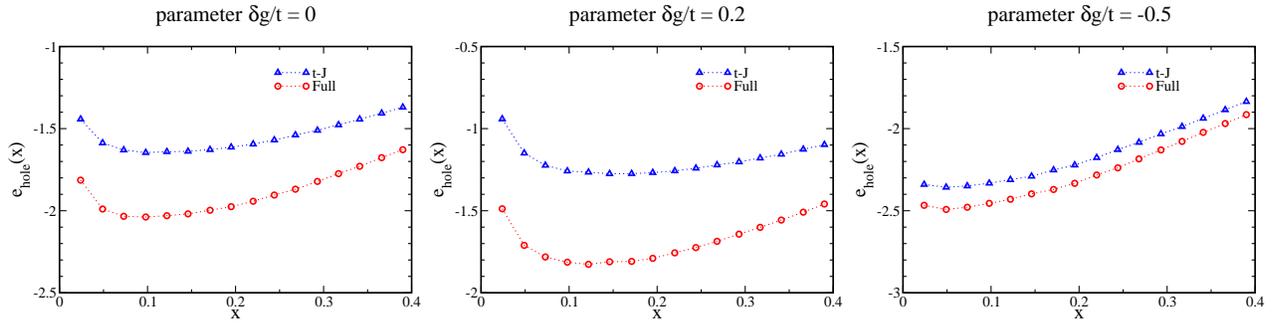

\includegraphics[angle=0,width=5.5cm,clip]{ehole_dg0.eps}
\includegraphics[angle=0,width=5.5cm,clip]{ehole_dg02.eps}
\includegraphics[angle=0,width=5.5cm,clip]{ehole_dgm05.eps}
\caption{(Color online)
Comparison of the energy density per hole $e_{hole}(x)$ for the {\it full} and the {\it t-J} Hamiltonians at different values of  $\delta g/t$: $\delta g/t = -0.5$ ($U/t = 53.32$), $0$ ($U/t = U_o/t = 13.33$) and $0.2$ ($U/t = 9.26$). For the {\it full} Hamiltonian, the minimum of $e_{hole}(x)$ is more clearly defined signaling 
greater tendency for phase separation. The unit of energy is $t$.}
\label{fig_1b}
\end{figure*}

\begin{figure*}[]
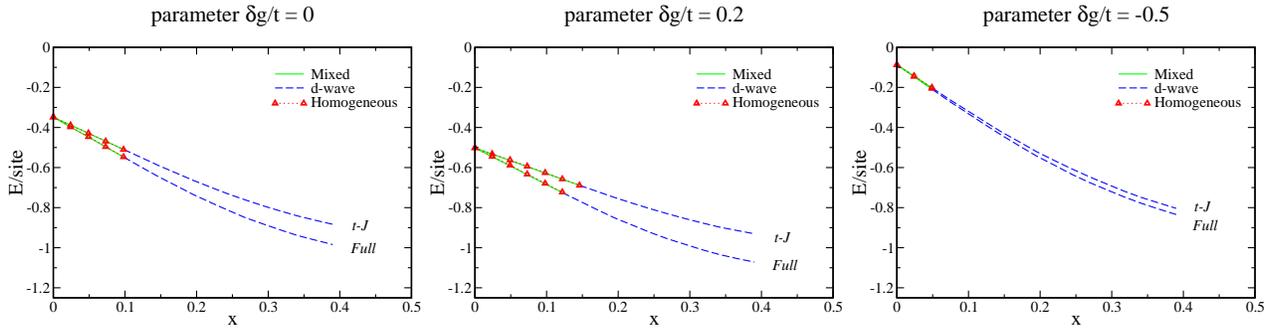

\includegraphics[angle=0,width=5.5cm,clip]{homcomp_dg0.eps}
\includegraphics[angle=0,width=5.5cm,clip]{homcomp_dg02.eps}
\includegraphics[angle=0,width=5.5cm,clip]{homcomp_dgm05.eps}
\caption{(Color online)
Comparison of the ground state energies for the {\it full} and the {\it t-J} Hamiltonians at different
values of  $\delta g/t$: $\delta g/t = -0.5$ ($U/t = 53.32$), $0$ ($U/t = U_o/t = 13.33$) and $0.2$ ($U/t = 9.26$).
The {\it full} Hamiltonian includes three-site hopping terms (Eq. \ref{eqn_heff}) that lower the energy relative to the {\it t-J} model. At zero hole doping, three-site terms are suppressed and both models converge.
The unit of energy is $t$.}
\label{fig_2b}
\end{figure*}

\begin{figure}[]
\includegraphics[angle=0,width=5cm,clip]{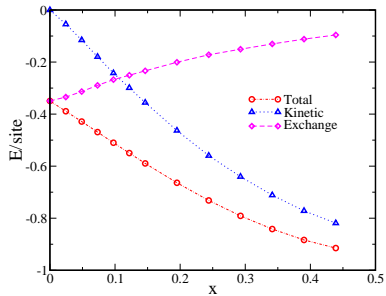}
\caption{(Color online) Doping dependence of the total, kinetic and exchange energies for the {\it t-J} model.
The unit of energy is $t$. } 
\label{fig_3}
\end{figure}

In the Figs. \ref{fig_1b} and \ref{fig_2b}, a comparison of the {\it full} and the {\it t-J} Hamiltonians at different values of $\delta g/t$
 is given. Both models converge at the zero doping $x=0$ since three-site hopping terms are density suppressed. 
However, at finite doping, differences arise: as seen in the Fig. \ref{fig_1b}, the $e_{hole}(x)$ tends to have a slightly larger curvature leading 
to a clear definition of $x_c$. This tendency is more noticeable for $\delta g/t> 0$ ($U < U_o$). But, for $U << U_o$, the validity of the perturbative Hamiltonian (Eq. \ref{eqn_eff0} and \ref{eqn_heff}) is being compromised.
 Here, the critical $x_c$ seems to be identical for these two models. 

\section{CONCLUDING REMARKS} 

 We have investigated the appearance of different quantum phases
in a system of strongly interacting fermions in the two dimensional optical lattices. We have shown that
the interaction strength $U$ has a direct impact on the shape of the phase diagram. 
In the laboratory, the optical lattice is usually placed within a smoothly varying harmonic trap.
Then, the usual shell structure is expected to appear. We have shown that the
mixed phase is energetically allowed and possibly more robust than the homogeneous phase.

There are several experimental techniques to measure the quantum correlations: 
for example, radio frequency techniques to directly measure the excitation spectrum of the Fermi gas are being developed (see Ref. \cite{jin08}).
 The measurement of the correlations in the noise could also be used, as it was applied to the detection of the Mott insulator phase (see Ref. \cite{bloch08}). 
Also, Bragg scattering is a scheme that allows the detection of the nodal points in the momentum space \cite{hofstetter02}.
A connection of the single band model to the resonant regime was made through perturbation theory,
 although a more elaborate multi-band model might be required for improved description. 

We acknowledge support by DARPA BAA 06-19. SP thanks Department of Science and Technology (DST) for funding 
and APS-IUSSTF Physics Student Visitation Program for travel support. SYC and NT also acknowledge support from ARO W911NF-08-1-0338 and NSF-DMR 0706203. We thank P. Zoller, M. Baranov, and U. Schneider for helpful discussions.

\section{APPENDIX} 

 For the resonant fermions with $k_Fa$ large,
multi-band effects have to be included. A model by Duan\cite{duan05,duan08}
accounts them by an effective single-band Hubbard-type model. This approach
physically corresponds to taking into account only the lowest energy but
exact one-particle and two-particle states on each lattice site. For a
single-particle state, the corresponding wave function is simply the ground
sate wave function in the local potential well. In the strongly interacting
limit, the wave function of a two-particle state includes contributions of
all excited states and does not reduce to a properly symmetrized product of
the two ground state wave functions. As a result, the overlap of the wave
functions for nearest neighbor sites and, therefore, the hopping amplitude
between them will depend on a particular density configuration (see Fig. \ref{fig_OL}): The hopping
amplitude $t_{01}= t_{10} = t$ for the hop from a singly occupied site to an empty
site, will be different from the hopping amplitudes $t_{11}=t_{02}=t_{20}=t+\delta g
$ for a hop from a singly occupied site to another singly occupied one and for a
reverse hop from a doubly occupied site to an empty one, respectively. This
is because the amplitude $t_{01}$ contains the overlap between the two
ground state wave functions on the nearest-neighbor sites, while the
amplitudes $t_{11}=t_{02}$ are determined by the overlap of the \textsl{exact%
} two-particle wave function on a site with the product of the ground state
wave functions on the same and\ the nearest-neighbor sites. In turn, the
hopping amplitude $t_{12}=t_{21}=t+\delta t$ for a hop from a doubly occupied site
to a singly occupied site contains the overlap of the exact two-particle
wave functions on nearest-neighbor sites and, hence, will be different from
both $t_{01}$ and $t_{11}=t_{02}$. As a result, the Hamiltonian of the
extended Hubbard model reads 
\begin{eqnarray}
\mathcal{H} &=&-\sum\limits_{<i,j>\sigma }[t+\delta g(n_{i,\bar{\sigma}%
}+n_{j,\bar{\sigma}})+\delta t(n_{i,\bar{\sigma}}n_{j,\bar{\sigma}%
})]c_{i,\sigma }^{\dagger }c_{j,\sigma }  \nonumber \\
&&+U_o\sum\limits_{i}n_{i,+1}n_{i,-1},  \label{eqn_hamil_duan}
\end{eqnarray}%
where the density assisted hopping terms in the kinetic energy is
characterized by $\delta g$ and $\delta t$. 

The coefficients $\delta g/t$ and $\delta t/t$ depend on the characteristics
of the atomic species such as the width (in magnetic field) of the resonance 
$\Delta B$, the background scattering length $a_{bg}$ and the difference of
the atomic magnetic moments between the closed and the open scattering
channels $\Delta \mu $. In particular, $\delta g/t$ can be written as 
$\delta g/t\approx (0.38(V_{o}/E_{r})^{0.25}-1)+\alpha $ (see the Fig. 1 of
the Ref. \cite{duan05}). $\alpha $ is a weakly dependent quantity of 
$V_{o}/E_{r}$ and proportional to $\sqrt{\Delta B|a_{bg}|\Delta \mu }$. Thus,
for a given atomic species and a resonant magnetic field, $\delta g/t$ can
be tuned over a range of the optical lattice depths. 

We define the kinetic
operators $\mathcal{K}_{m}$ as 
\begin{eqnarray}
\mathcal{K}_{0} &=&-t\sum\limits_{<i,j>\sigma }h_{i,\bar{\sigma}}c_{i,\sigma
}^{\dagger }c_{j,\sigma }h_{j,\bar{\sigma}}  \nonumber \\
&&-(t+2\delta g+\delta t)\sum\limits_{<i,j>\sigma }n_{i,\bar{\sigma}}c_{i,\sigma }^{\dagger }c_{j,\sigma }n_{j,\bar{\sigma}}  \nonumber \\
\mathcal{K}_{1} &=&-(t+\delta g)\sum\limits_{<i,j>\sigma }n_{i,\bar{\sigma}}c_{i,\sigma }^{\dagger }c_{j,\sigma }h_{j,\bar{\sigma}}  \nonumber \\
\mathcal{K}_{-1} &=&-(t+\delta g)\sum\limits_{<i,j>\sigma }h_{i,\bar{\sigma}}c_{i,\sigma }^{\dagger }c_{j,\sigma }n_{j,\bar{\sigma}}
\end{eqnarray}%
 $\mathcal{K}_{m}$ change the number of doubly occupied sites by $m$ and 
obeys commutation relation $[\mathcal{V},\mathcal{K}_{m}]=mU_o\mathcal{K}_{m}$ with $\mathcal{V}=U_o\sum\limits_{i}n_{i,+1}n_{i,-1}$) 

Here, we have used the usual definition $h_{i,\sigma }\equiv 1-n_{i,\sigma }$. The \textit{full} Gutzwiller projector can be shown to drop the interaction and the $\delta t$ dependent terms. The effective Hamiltonian after
Schrieffer-Wolff transformation is then the same as Eq. \ref{eqn_heff} with the definition of the effective $U\equiv U_o/(1+\delta g/t)^{2}$. In a recent work \cite{wang08} on the one dimensional attractive fermionic gas with population imbalance, it was found that the pairing order gets
enhanced with $\delta g<0$, while the spin-spin correlation is suppressed.
Obviously, due to the difference in the dimensionality (1D vs 2D) and the nature
of the interaction (attractive vs repulsive), an intuitive connection
is rather complicated.

\end{document}